\documentclass[%
 reprint,
 amsmath,amssymb,
 aps,
 prl,
 superscriptaddress,
]{revtex4-1}

\usepackage{graphicx}
\usepackage{bm}
\usepackage{color}
\usepackage{amsmath}
\usepackage{graphicx}
\usepackage[section]{placeins}
\usepackage{amsmath}

\usepackage[
margin=0.71in,
]{geometry}

\begin{document}
\preprint{APS/123-QED}
\title{Free-space remote sensing of rotation at photon-counting level}%

\author{Wuhong Zhang}
\author{Jingsong Gao}
\author{Dongkai Zhang}
\author{Yilin He}
\author{Tianzhe Xu}
\affiliation{Department of Physics, Jiujiang Research Institute and Collaborative Innovation Center for Optoelectronic Semiconductors and Efficient Devices, Xiamen University, Xiamen 361005, China}
\author{Robert Fickler}
\email{rfickler@uottawa.ca}
\affiliation{Department of Physics, University of Ottawa, 25 Templeton St., Ottawa, Ontario K1N 6N5, Canada}
\author{Lixiang Chen}
\email{chenlx@xmu.edu.cn}
\affiliation{Department of Physics, Jiujiang Research Institute and Collaborative Innovation Center for Optoelectronic Semiconductors and Efficient Devices, Xiamen University, Xiamen 361005, China}

\date{\today}

\begin{abstract}
The rotational Doppler effect associated with light's orbital angular momentum (OAM) has been found as a powerful tool to detect rotating bodies. However, this method was only demonstrated experimentally on the laboratory scale under well controlled conditions so far. And its real potential lies at the practical applications in the field of remote sensing. We have established a 120-meter long free-space link between the rooftops of two buildings and show that both the rotation speed and the rotational symmetry of objects can be identified from the detected rotational Doppler frequency shift signal at photon count level. Effects of possible slight misalignments and atmospheric turbulences are quantitatively analyzed in terms of mode power spreading to the adjacent modes as well as the transfer of rotational frequency shifts. Moreover, our results demonstrate that with the preknowledge of the object's rotational symmetry one may always deduce the rotation speed no matter how strong the coupling to neighboring modes is. Without any information of the rotating object, the deduction of the object's symmetry and rotational speed may still be obtained as long as the mode spreading efficiency does not exceed $50~\%$. Our work supports the feasibility of a practical sensor to remotely detect both the speed and symmetry of rotating bodies.
\end{abstract}

\maketitle
\section{I. INTRODUCTION}
The Doppler effect is a well-known phenomenon describing the frequency shift of a wave, such as sound waves and light waves. The frequency emitted by a moving source becomes higher as the source is approaching an observer, while it becomes lower as it is receding. This linear Doppler effect is widely used for sonar and radar systems to deduce the speed of a moving object \cite{raymond1984laser}. In contrast to linear motion, a rarely encountered example is the angular version of the Doppler shift arising from rotation \cite{padgett2006electromagnetism}. The rotational Doppler effect was first observed by Garetz and Arnold, who used rotating half-wave plates of angular velocity $\Omega$  to imprint a frequency shift $2\Omega$ to circularly polarized light \cite{garetz1979variable}. Such a frequency shift is in essence associated with spin angular momentum of photons, and can be understood from the dynamically evolving geometric phase in light of the Poincare sphere \cite{simon1988evolving}. More recently, the rotational Doppler effect was also verified in the second-harmonic generation with a spinning nonlinear crystal of three-fold rotational symmetry \cite{li2016rotational}. In addition to spin, a light beam with a twisted phase front of $\exp(i\ell\phi)$ carries $\ell\hbar$ orbital angular momentum (OAM) per photon, where $\phi$ is the azimuthal angle and  $\ell$ is an integer \cite{allen1992orbital}.It was first demonstrated by Courtial et al. that a rotating Dove prism could impart the OAM beams with a frequency shift of $\ell\Omega$, where $\Omega$  denotes the angular velocity \cite{courtial1998measurement,courtial1998rotational}. Via the coherent interaction of OAM beams with atom samples, Barreiro et al. reported on the first spectroscopic observation of rotational Doppler shift \cite{barreiro2006spectroscopic}. Also, Korech et al. developed an intuitive method to observe the molecular spinning based on the rotational Doppler effect \cite{korech2013observing}.

Recently, considerable attention was paid to exploit the twisted light's rotational Doppler Effect to detect rotating bodies. Lavery et al. demonstrated that the angular speed can be deduced by detecting the frequency shift of on-axis OAM components that are scattered from a spinning object with an optically rough surface \cite{lavery2013detection}. They further employed an OAM-carrying white-light beam and only observed a single frequency shift within the same detection mode \cite{lavery2014observation}. Rosales-Guzm¨¢n et al. showed that under different OAM-mode illumination the full three-dimensional movement of particles could be characterized by using both the translational and rotational Doppler Effect \cite{rosales2014measuring}. Fang et al. experimentally demonstrated that both the rotational and linear Doppler effect actually share a common origin such that one can use one effect to drive the other effect \cite{fang2017sharing}. Zhao et al. extended the OAM illumination beam into the radio frequency area and detected the speed of a rotor in a proof-of-concept experiment \cite{zhao2016measurement}. Based on the rotational Doppler Effect, Zhou et al. devised an OAM-spectrum analyzer that enables simultaneous measurements of the power and phase distributions of OAM modes \cite{zhou2017orbital}. Although the real potential of the rotational Doppler effect lies at its practical application in the non-contact remote sensing \cite{padgett2014new,marrucci2013spinning}, we note that thus far no experimental verification of rotational Doppler effect has been conducted outside of the laboratory.

The challenges for a long-distance implementation originate from the OAM mode spreading induced by atmospheric turbulence and the low photon collection efficiency due to beam divergence misalignment \cite{willner2015optical}. Here we report observation of the rotational Doppler effect over a 120-meter free-space link between the rooftops of two buildings across the Haiyun Campus of Xiamen University in a city environment. Our scheme works with extremely weak light illumination by employing a single-photon counting module. This work moves a step towards the remote sensing applications in realistic environment with twisted light's rotational Doppler effect.

\section{II. THEORETICAL ANALYSIS AND SIMULATIONS}
We assume that the rotating object is mathematically described by a complex function $\psi(r,\varphi)$ in the cylindrical coordinates and is illuminated with a fundamental Gaussian mode. As the  Laguerre Gaussian(LG) modes form a complete and orthogonal basis, we can describe the light field reflected from the object as $\psi(r,\varphi)=\sum\nolimits_{\ell,p}{A_{\ell ,p}}\rm{LG}_p^\ell(r,\varphi )$, where ${\rm{LG}}_p^\ell (r,\varphi )$ denotes the LG mode with azimuthal and radial indices $\ell$ and $p$, respectively, and ${A_{\ell ,p}} = \int{\int{{{[{\rm{LG}_p^\ell( {r,\varphi})}]}^*}\psi({r,\varphi})rdrd\varphi}}$ is the overlap amplitude. The method of using a coherent superposition of LG modes to represent an object was called digital spiral imaging by Torner et al. \cite{torner2005digital}, and has been found as an effective technique to encode optical information and retrieve topographic information of an object, particularly useful for objects of high spatial symmetry \cite{molina2007probing,petrov2012characterization,chen2014quantum}. We plot the LG mode spectra, $|A_{\ell,p}|$, for two typical objects, i.e., a three-leaf Clover and a five-petal Pentas in Fig. 1. One can see from the top panel of Fig. 1, the dominant LG modes are those with the azimuthal indices of  $\ell=0, \pm3, \pm6, \cdots$ (Fig. 1a) and  $\ell=0, \pm5, \pm10, \cdots$ (Fig. 1b), being associated with the three-fold and five-fold rotational symmetry of the Clover and Pentas, respectively. This symmetry can be illustrated more evidently by the pure OAM spectra characterized by $P_\ell=\sum_p|A_{\ell,p}|^2$, as a sum of the mode weights over the radial $p$ index. In the lower panel of Fig. 1, we plot the pure OAM spectra for the three-leaf Clover (Fig. 1c) and five-petal Pentas (Fig. 1d). This follows closer the experimental expectation, because our detection technique uses a pure phase grating that only distinguishes between different $\ell$ modes irrespective of their radial structure. More details of the mode expansion can be found in our recent paper \cite{zhang2016encoding}.
\begin{figure}[t]
\centerline{\includegraphics[width=\columnwidth]{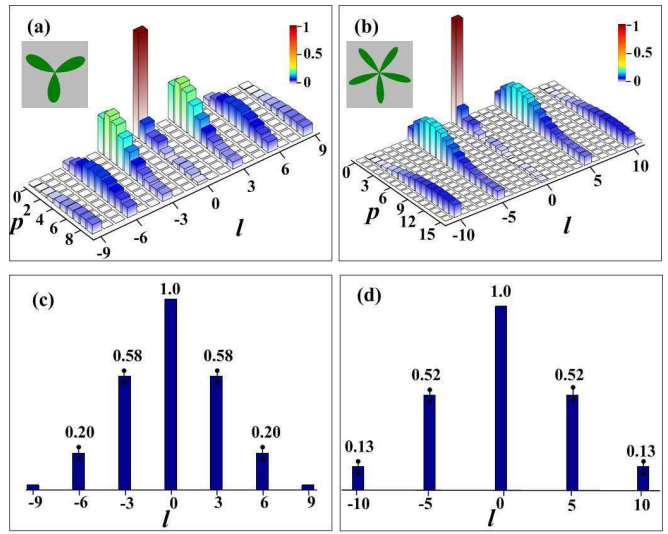}}
\caption{Spiral spectrum of the image gives an intuitive indication of the object symmetry.  Top panel is the peak-normalized LG mode spectra while bottom is the pure OAM spectra (a), (c): a three-leaf Clover. (b), (d): a five-petal Pentas.}
\end{figure}

The frame work of digital spiral imaging provides an intuitive understanding of the mechanism of the rotational Doppler effect. If the object is rotated by a constant angular speed $\Omega$  around its own axis, a time-varying phase shift of  $\ell\Omega t$ will be imparted to each OAM eigenmode. As a consequence, we rewrite the transmitted or reflected light fields as
\begin{eqnarray}
\psi(r,\varphi ,t)=\sum\limits_{\ell,p}{{A_{\ell ,p}}{\rm{LG}}_p^\ell(r,\varphi )\exp(i\ell \Omega t)} ,
\end{eqnarray}
which manifests the OAM-dependent frequency shift $\ell\Omega$. By using specific OAM superposition modes $\Phi({r,\varphi})=[\exp(i\ell\phi)+\exp(-i\ell\phi)]/\sqrt 2 $  to detect the reflected light fields, we can obtain the signal as
\begin{eqnarray}
I(t)&=&\left|\int\int\Phi^*(r,\varphi)\psi(r,\varphi,t)rdrd\Phi\right|^2 \nonumber \\
&&\propto 2P_\ell[\cos(2\ell\Omega t)+1]
\end{eqnarray}
assuming $P_\ell=P_{-\ell}$ . Hence, we find an intensity modulation of frequency $f_{mod}=\frac{|2\ell \Omega|}{2\pi}$ for the detected superposition modes $\Phi({r,\varphi})$. The maximum detected intensity modulation is only determined by the mode spectrum $P_\ell$ of the reflected light fields. It is obvious that one cannot detect the intensity modulation signal with the OAM mode spectrum $|P_\ell|=0$. By scanning the intensity modulation signal of the entire OAM spectrum, the rotational symmetry of object may be also deduced. Generally, as all of the constituent LG modes in Eq. (1) propagate individually in free space without coupling with each other, the mode weights $A_{\ell,p}$, i.e. the OAM spectra, will remain unchanged without any influence by atmospheric disturbance \cite{zhang2016experimental}. This indicates that the measurement of rotational Doppler effect $f_{mod}$ at the distant receiver will be equivalent to that at the transmitter.

\section{III. EXPERIMENTAL SETUP}
We implement the 120-m free-space link between the rooftops of two buildings. Fig. 2 shows the transmitter and receiver locations. At transmitter (top-left inset in Fig. 2), the fundamental Gaussian mode from a 633 nm Helium-Neon (HeNe) laser beam (Thorlabs HNL210L), after being collimated by the first telescope, illuminates a computer-controlled spatial light modulator (SLM). We prepare the desired holographic gratings with the SLM to mimic the rotating objects, e.g., a Colver and a Pentas. The first diffraction order, which is selected by a 4-f filtering system ($f_1$=500 mm, $f_2$=150 mm), then acquires the profile of the rotating objects. The beam diameter on the image plane is about 2 mm, which is the diffraction-limited input beam diameter of the second telescope (Thorlabs GBE20-A). The telescope is used to further expand the beam to a diameter of about 40 mm, and is followed by the transmission of the rotating pattern over the 120-m free-space link to the receiver. Due to diffraction of the image, we get a beam diameter about 50 mm at the receiver. There, another 4-f filtering system with a collection lens of 100 mm in diameter ($f_3$=500 mm) and a re-imaging lens ($f_4$=75 mm) is used to demagnify the collected beam to have a diameter of $\sim7.5$ mm. Considering the limited aperture of the lens, we estimate that approximately $95~\%$ of the incoming light is collected to illuminate the second SLM. A holographic grating with desired superposition of $\pm\ell'$  OAM modes is displayed on another SLM and together with a third 4-f filtering system is used to select the first diffraction order and image the SLM plane on the single mode fiber. The grating flattens the phase of one specific superposition mode. Because only modes with a plane phase front are coupled into a single mode fiber, this arrangement acts as a mode filter and we experimentally measure the overlap probability as described by Eq.(2).
\begin{figure}[t]
\centerline{\includegraphics[width=\columnwidth]{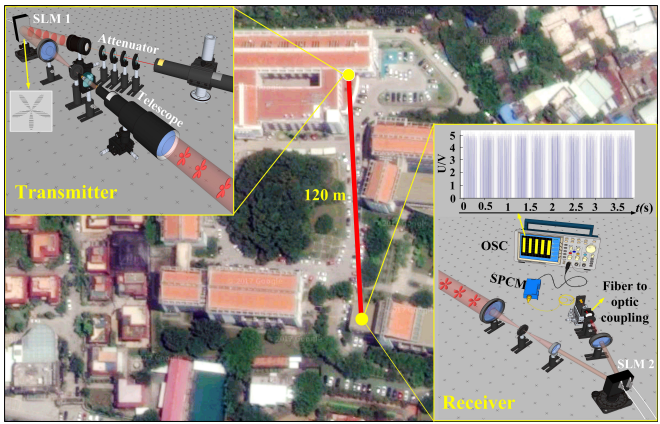}}
\caption{120-m free-space optical link implemented from building to building in the Haiyun Campus of Xiamen University. Left-top inset, optical setup of the transmitting terminal; Right-bottom inset, optical setup of the receiving terminal, SLM: spatial light modulator, OSC: Oscilloscope, SPCM: Single photon counting module.}
\end{figure}
In the experiment, we start with using the full power of the HeNe laser with 21 mW to align the whole optical path. Then we use a series of neutral-density filters to attenuate the laser beam to a very faint level and demonstrate the ability of our scheme to work at the photon-counting level. We employ the single photon counting module (SPCM, Excelitas) to detect single photons that are phase-flattened and coupled into the single-mode fiber. The single-photon events are monitored by a Digital Phosphor Oscilloscope (DPO3012, Tektronix) over the measurement time of a few minutes. An exemplary measured data set of single photon detections can be seen in the right-bottom inset of Fig. 2. Each blue line in the graph represents a detected photon, and the degree of sparsity represents the number of the detected photons varying with time. By subsequent applying a fast Fourier transform (FFT) to the time-varying photon count sequence, we extract directly the frequency shifts. It is noted that the optical alignment between the receiver and transmitter is very important, as both lateral displacement and angular deflection could cause severe modal crosstalk \cite{vasnetsov2005analysis,lavery2017free}. To minimize possible misalignments, we couple a second HeNe laser at the receiver into the fiber before each measurement and send the light back to the transmitter to make sure the forward and backward propagating beams are well overlapped at both stations. Moreover, to minimize vibrations induced by people walking around in the building of the sender, we performed all experiments after midnight.

\begin{figure}[t]
\centerline{\includegraphics[width=0.8\columnwidth]{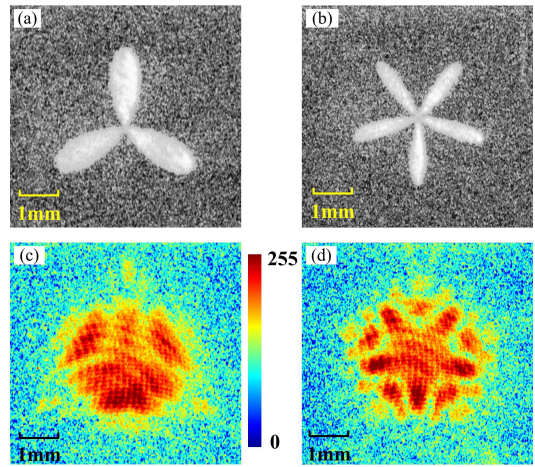}}
\caption{Weak-light images of the rotating objects. (a) and (b), show the gray scale images of  the Clover and Pentas recorded by EMCCD camera at transmitter, respectively. (c) and (d), show the false color images of the Clover and Pentas images at receiver, respectively.}
\end{figure}

\section{IV. EXPERIMENTAL RESULTS}

We first investigate the propagation features of the Clover and Pentas images transmitted through the 120-m free-space link. We use a low-noise electron multiplier CCD camera (EMCCD, E2V, $768\times576$ pixels) to record the light fields at both the transmitter and receiver. Fig. 3(a) and 3(b) display the Clover and Pentas images at the transmitter, respectively. After propagating through the 120-m free-space link we see in Figs. 3(c) and 3(d) that both the Clover and Pentas at the receiver aperture become barely recognizable, as a main consequence of the free-space diffraction. We estimate that the photon flux recorded in each of these images is $\sim10^5~\text{s}^{-1}$ based on the sender power of the light field. At the receiver, we use holographic gratings for measuring specific OAM superposition, with diffraction efficiency $\sim20~\%$. The fiber-optic coupling efficiency is $\sim80~\%$ and the SPCM detection efficiency is $\sim60~\%$ at 633 nm. Thus we record approximately $10^3~\text{s}^{-1}$ single-photon events.

\begin{figure}[t]
\centerline{\includegraphics[width=\columnwidth]{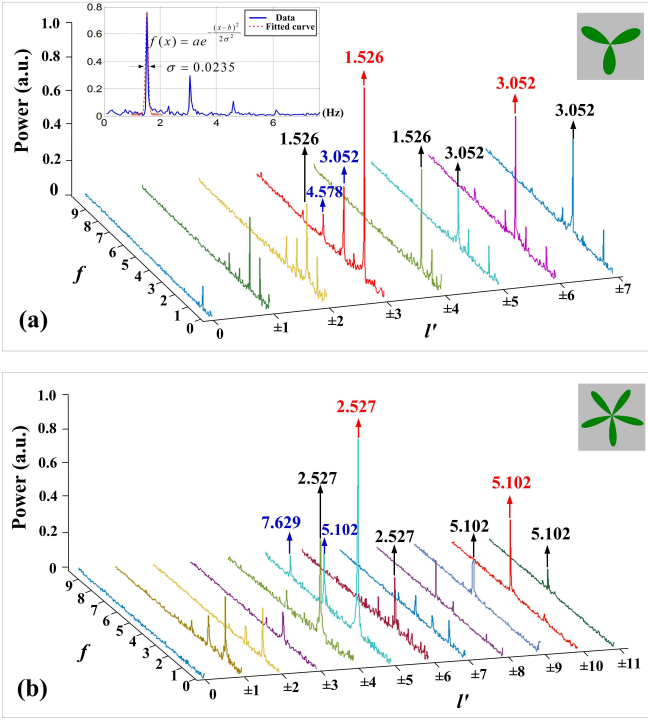}}
\caption{Measured power spectra of rotational Doppler shifts. a, Clover. b, Pentas. Both are set to rotate at the same angular rate ${\Omega_1} = 90^\circ~\text{s}^{-1}$. The obtained frequencies for each OAM mode are represented with different colors. The value of the frequency labeled in red text denotes the main mode¡¯s frequency, which are used to deduce the rotation speed. The inset of (a) denote the data process of one of the main modes $\ell'=\pm3$. The black texts denote the undesired frequency which is mainly caused by slight misalignment. The blue texts besides the main modes denote the additional frequency that might be caused by the slight deviation of photon counts from the standard sinusoid.}
\end{figure}
\begin{figure}[t]
\centerline{\includegraphics[width=\columnwidth]{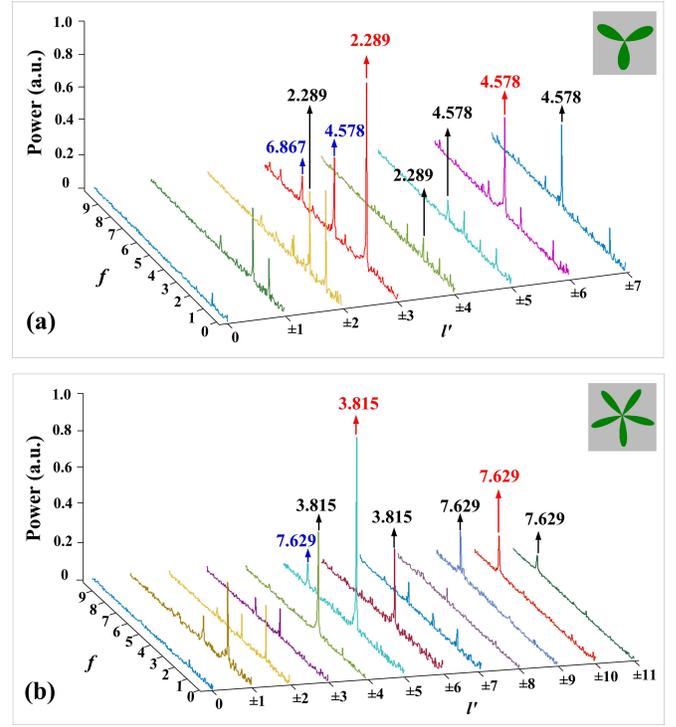}}
\caption{Measured power spectra of rotational Doppler shifts. a, the Clover. b, the Pentas. Both are set to rotate at the same angular rate ${\Omega_2}=135^\circ~\text{s}^{-1}$. The different colors are used for the same purpose as described in the caption of Fig. 4.}
\end{figure}

Without losing generality, we restrict our first set of measurements to the Clover and Pentas with a rotating velocity ${\Omega_1}=90^\circ~\text{s}^{-1}$  corresponding to a rotational frequency of 0.25 Hz. According to Eq. (2), we measure OAM superspositions ranging from  $\ell'=\pm1$ to $\ell'=\pm7$  for clover and up to  $\ell'=\pm11$ for Pentas, respectively. After applying a fast Fourier Transform (FFT) directly to the time-varying single-photon events, we obtain the experimental results in Fig. 4. The peak power frequency observed at $\ell'=\pm3$  with  $f=$1.526 Hz and  $\ell'=\pm6$ with $f=$3.052 Hz for the Clover in Fig. 4(a), while at $\ell'=\pm5$ with $f=$2.527 Hz and $\ell'=\pm10$ with $f=$5.102 Hz for the Pentas in Fig. 4(b), which manifests again the three-fold and five-fold rotational symmetry of the Clover and the Pentas. It is noted that the full width at half maximum (FWHM) of the peak frequency is very narrow, which shows the accuracy of our measurements. We use a Gaussian fitting curve to calculate the Gaussian RMS width $\sigma$ of each peak frequency, one example of $\sigma\approx0.0235$ for $\ell'=\pm3$  is shown in the inset of Fig. 4(a). According to $f_{\bmod}=|2\ell\Omega|/{2\pi}$, we deduce the rotating velocity with $\Omega_{Clover}=({91.44^\circ}\pm {1.4^\circ})~\text{s}^{- 1}$ and $\Omega_{Pentas}=({90.97^\circ}\pm{0.97^\circ})~\text{s}^{-1}$. We further set both the Clover and Pentas to rotate at ${\Omega_2}=135^\circ~\text{s}^{-1}$ corresponding to a rotational frequency of 0.375 Hz to test our setup. From the measured power spectra in Fig. 5(a), we find the peak frequency of $f=$2.289 Hz are detected in modes $\ell'=\pm3$, and $f=$4.578 Hz in modes $\ell'=\pm6$ for the rotating Clover. For the Pentas object we find peak frequencies $f=$3.815 Hz in modes $\ell'=\pm5$, and $f=$7.629 Hz in modes $\ell'=\pm10$ (see Fig. 5(b)). The analog data processing gives the deduced rotating velocity $\Omega _{Clover}=({137.34^\circ}\pm{1.32^\circ})~\text{s}^{-1}$ and $\Omega_{Pentas}=({137.34^\circ }\pm{1.18^\circ})~\text{s}^{-1}$. Hence, from the obtained results the rotation speed can be deduced with very high accuracy, which demonstrates the validity of our 120 m free space sensing of rotation at photon-counting level. Moreover, in the absence of any information about the detected rotation object, the peak power of the detected scanning OAM frequency spectrum also provided a way to analyze the object¡¯s rotational symmetry.

In addition to the expected signal from the rotating pattern, we also detected the frequencies $f=$3.052 Hz, 4.578 Hz in the main mode $\ell'=\pm3$ for the Clover and $f=$5.102 Hz, 7.629 Hz in $\ell'=\pm5$ for the Pentas under a rotation speed ${\Omega_1}=90^\circ~\text{s}^{-1}$, as denoted by the blue text in Fig. 4(a) and Fig. 4(b). Similarly, under a different rotation speed ${\Omega_2}=135^\circ~\text{s}^{-1}$, the same effect can be  observed with $f=$4.578 Hz, 6.867 Hz  in the main mode $\ell'=\pm3$  for the Clover and $f=$7.629 Hz in $\ell'=\pm5$  for the Pentas, as the blue text labeled in Fig. 5(a) and Fig. 5(b), respectively. According to Eq.(2), the photon counts over time should appear like a sinusoidal oscillation with $\ell'=\pm3$ and $\ell'=\pm5$, and the corresponding Fourier transform should only give one peak, respectively. However, in our practical detection the signal was obtained over a few minutes. Over that time frame, any slight misalignment caused by tiny vibrations of the transmitter, leads to a deviation from the expected sinusoidal recording. Such slight deviations from the standard sinusoid of photon counts will cause the higher harmonic peaks of frequency shifts.

As mentioned before, we should observe no modulation of the counts, i.e. $f=$0 for those OAM modes with $|P_\ell|=0$ as the simulation results in Fig. 1(c) and 1(d). However, in contrast to this theoretical expectation we also find the frequency $f=$1.526 Hz, $f=$2.289 Hz in the adjacent modes with $\ell'=\pm2,\pm4$, and $f=$3.052 Hz, $f=$4.578 Hz in $\ell'=\pm5,\pm7$ for the different rotation frequency of Clover object, as labeled by the black texts in Fig. 4(a) and 5(a). This effect is more dominant for Pentas with different rotation speed: $f=$2.527 Hz, $f=$3.815 Hz in modes $\ell'=\pm4,\pm6$, and$f=$5.102 Hz, $f=$7.629 Hz in modes $\ell'=\pm9,\pm11$. We attribute this effect to energy spreading accompanied with the transfer of frequency shifts from the dominant modes to the adjacent ones. If we have preknowledge of an object's symmetry, this could be identified as erroneous measured frequency shifts such that the rotation speed will still be simply deduced from the dominant modes. However, if there is no information about the object, one needs to measure different OAM components $\ell'$ and deduce the symmetry from the measured spectrum to determine the dominant mode. In this case, if we measure $\ell'$, it is possible to obtain a mixing of frequency shifts from different OAM modes $\Delta\omega=\{2\ell'_1\omega,2\ell'_2\omega,2\ell'_3\omega,\cdots\}$. One reasonable assumption might be that with an energy spread to adjacent modes of more than $50~\%$ a correct discrimination of the main mode is not easily possible anymore, such that the real frequency shifts and the symmetry of the object cannot be accurately deduced. Hence, the main contributions to the mode spreading phenomenon will be carefully analyzed in the following paragraphs.

Firstly, it might be caused by slight misalignments between the grating and detection pattern in the experiment. Lavery et al. have demonstrated that the desired LG mode would expand into its adjacent mode with about $25~\%-35~\%$ when there is a lateral displacement of $\Delta{x_0}=0.5{w_0}$, or a tilt angle of $\Delta\alpha=0.5\lambda/{w_0}$ between the grating and the detection pattern in lab scale \cite{lavery2011measurement}. In our experimental setting a small tilt angle vibration $\Delta \theta$ will cause a magnified displacement $\Delta d=\tan(\Delta \theta)\cdot L\approx\Delta\theta\cdot L$ between the received pattern and the grating, where $L$ is the distance between the sender and receiver. In this case, our detected signal should have an overlap probability:
\begin{eqnarray}
I(t)&=&\left|\int\int\Phi^*(x,y)\psi(x+\Delta d,y+\Delta d,t)dxdy\right|^2 \nonumber \\
&&\propto \left|{\sum\limits_\ell{\left( {{B_{\ell',\ell }}+{B_{-\ell',\ell }}}\right)}\exp(i\ell\Omega t)} \right|^2,
\end{eqnarray}
\begin{figure}[t]
\centerline{\includegraphics[width=\columnwidth]{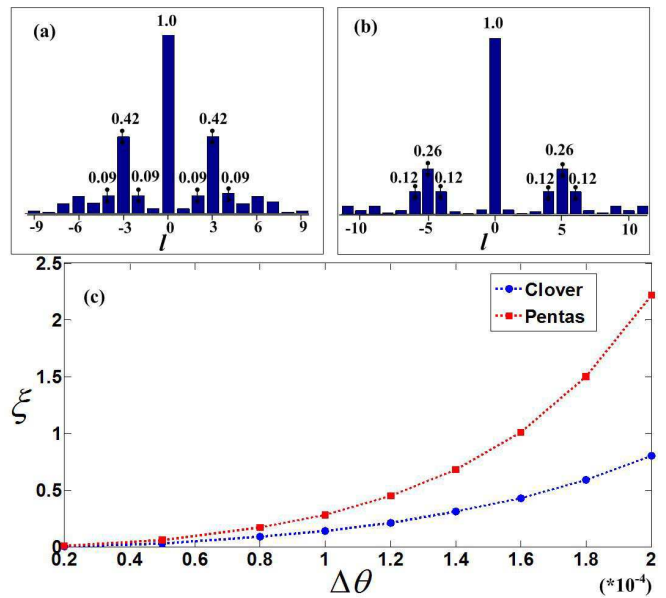}}
\caption{Influence of the misalignment on the mode spectrum. Simulation results of mode spreading effect when there is a small misalignment between the receiver pattern and the grating with $\Delta d=0.2w_0$ for (a): Clover object, (b): Pentas object. (c) The relationship of the spreading efficiency and the vibration angle in sender for both three-leaf Clover object and five-leaf Pentas object. Here the spreading efficiency is defined as $\xi_{clover}=(P'_{|\ell|=2}+P'_{|\ell|=4})/2P'_{|\ell|=3}$ for the Clover, while $\xi_{pentas}=(P'_{|\ell|=4}+P'_{|\ell|=6})/2P'_{|\ell|=5}$ for the Pentas}
\end{figure}
Note that we describe the beam in Cartesian coordinates here, as this is more convenient in our misalignment considerations. $B_{\ell',\ell}=\sum_{p',p}\int\int[{\rm{LG}}_{p'}^{\ell'}(x,y)]^{*}{\rm{LG}}_p^{\ell}(x,y)dxdy$   denotes the total coupling efficiency from the OAM mode $\ell$  to $\ell'$ due to the misalignment of the receiver pattern and the grating. If there is no mode spreading, i.e., $B_{\ell',\ell }=\delta _{\ell',\ell}$, we have $I(t)\propto {\left|{{A_{\ell '}}\exp(i\ell'\Omega t)+{A_{-\ell '}}\exp(-i\ell'\Omega t)}\right|^2}$, which is the trivial case without misalignment. In our experiment, the beam waist $w_0$ on the grating is about 7.5 mm. As an exemplary simulation to show the mode spreading of Clover and Pentas light field, we plot $P'_{\ell'}=\sum_\ell{\left|{{B_{\ell',\ell}}}\right|}^2$ in Fig. 6(a) and 6(b) with $\Delta d=0.2{w_0}$, which is corresponding to a small tilt angle vibration $\Delta\theta\approx{0.000013^\circ}$ at the sender. Such tiny swing angles are very likely caused by slight misalignments of the telescope, the SLM or the laser due to mechanical instabilities. To focus on the main modes coupling to the adjacent ones contributions we define a spreading efficiency as $\xi_{clover}=(P'_{|\ell|=2}+P'_{|\ell|=4})/2P'_{|\ell|=3}$ for the Clover and $\xi_{pentas}=(P'_{|\ell|=4}+P'_{|\ell|=6})/2P'_{|\ell|=5}$ for the Pentas, respectively. One can see (Fig. 6(a)) the dominant modes of $\ell'=\pm3$ are spreading to the modes $\ell'=\pm2$ and $\ell'=\pm4$ with an efficiency nearly $24~\%$. This effect is even stronger for the Pentas pattern, where a spreading to the next neighbouring modes of up to $48~\%$ can be observed (see Fig 6(b)). It seems that more complex objects lead to higher mode spreading effect. This observation becomes clearer, when we plot the spreading efficiency with respect to the vibration angle $\Delta\theta$ at sender (see Fig. 6(c)). Under the same vibration angle, the five-leaf object always has a higher spreading efficiency than the three-leaf object. From this investigations we see that minimizing vibrations at the sender are crucial, especially when longer measurement times are needed.

Secondly, for a practical free-space optical link, the atmospheric turbulence can lead to random variations in the refractive index such that the phase front of a propagating light is inevitably distorted \cite{paterson2005atmospheric}. This is particularly important for the detection of rotational Doppler effect, as its measurement is very sensitive to the optical filtering of suitable OAM superposition \cite{lavery2013detection}. Here we adopt the model developed by von Karman to describe the influence of turbulence \cite{lane1992simulation,ostashev1998coherence} and represent the atmospheric turbulence link as several turbulent phase screens, each separated by some distance of propagation \cite{fu2016influences,zhao2012aberration}. After propagating through the turbulence with a distance $Z$, the modified light field at the receiver can be written as, $\psi'(r,\varphi,Z,t)$. We first perform a numerical simulation with $Z=120m$ for the Clover and Pentas image under different air turbulence strength $C_n^2$  in Fig. 7(a). For comparison, we also show the experimental light field captured at receiver, as shown in Fig. 7(b). One can see the most deformation of the image comes from the diffraction but not the air turbulence. So it is reasonable to say that our air turbulence strength should on the order of $10^{-15}-10^{-14}$. To further simulate the effect of atmosphere turbulence on the rotational light field, we use the similar mode expansion method described earlier:
${A'_{\ell ,p}}=\int\int[{\rm {LG}_p^\ell(r,\varphi,Z)]}^*\psi'({r,\varphi,Z,t})rdrd\varphi$. When taking the effect of atmospheric turbulence with such a strength of $C_n^2=7.5*10^{-15}$  into account, we only find a very small influence on the pure OAM spectra characterized by $P'_\ell=\sum_p{\left|{A'}_{\ell ,p}\right|}^2$ (see Fig. 7(c)). For the Clover, the power of the dominant single OAM modes, i.e., $\ell'=\pm3$, is slightly spread to the adjacent modes with around $1~\%$ efficiency. For the five-leaf Pentas and the dominant modes $\ell'=\pm5$ to the adjacent modes with around $1.2~\%$. Again, this demonstrates that the more complex the pattern the more severe the mode coupling.

To see under which turbulence conditions the 120m link might not have worked, we do a further simulation and investigate the relationship between the strength of turbulence($C_n^2$) and the mode spreading efficiency($\xi$) (see Fig. 7(d)). Only strong turbulence (stronger than $5*10^{-14}$) will cause $\xi>50~\%$ and may cause the incorrect discrimination of the main mode from the adjacent one. Thus we believe that in our situation the atmospheric turbulence only contributed very weakly to the mode spreading phenomena observed in our results. This can be clearly seen in the frequency spectrum in Fig. 5(a), where nearly no signal can be detected for the modes $\ell'=\pm4$ and $\ell'=\pm5$, while during another measurement run (shown in Fig 4(a)) we observe stronger coupling to neighboring modes. This asymmetric phenomenon gives additional evidence that the mode spreading is mainly coming from the misalignment but not the turbulence.
\begin{figure}[t]
\centerline{\includegraphics[width=\columnwidth]{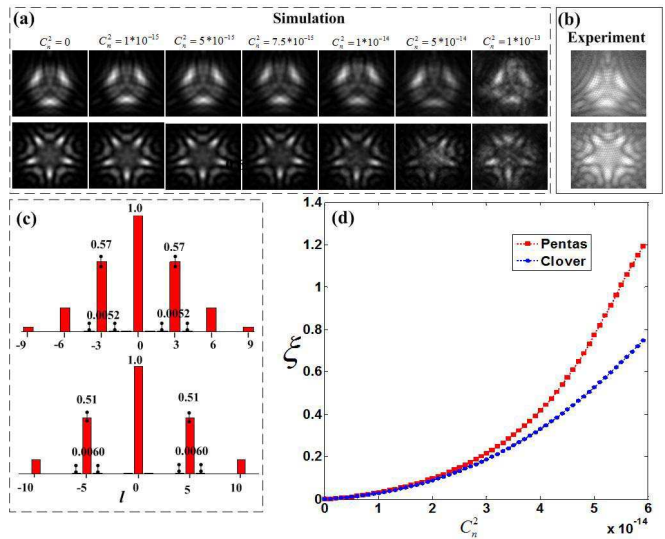}}
\caption{Influence of the atmospheric turbulence on received patterns. (a): simulation results of the Clover and Pentas light fields at the receiver under different turbulence strength. (b): experimentally obtained results recorded by an EMCCD at the receiver. (c): simulation results of the slightly disturbed pure OAM spectra with a turbulence strength of $C_n^2=7.5*10^{-15}$, which we estimated to resemble our experimental conditions. (d): simulation results of the variation of spreading efficiency with respect to different turbulence strength over 120 m. }
\end{figure}

However, irrespective where this mode spreading is coming from, misalignment or turbulence, one may always deduce the rotation speed and the rotational symmetry of an unknown rotating object as long as one can clearly discriminate the dominant modes from the spread modes. For a longer distance sensing of an rotating object, one will need to address both of the analyzed effects. Here, adaptive optics and machining learning-based pattern recognition may be introduced to compensate the effect of turbulence or misalignment\cite{ren2014adaptive}.

\section{V. CONCLUSION}

In conclusion, we have conducted an outdoor experiment of measuring the rotational Doppler effect by building a 120-m free-space optical link in a realistic city environment. Our experimental results with two typical rotating objects, i.e., Clover and Pentas patterns, demonstrate that long-distance remote sensing of spinning bodies is practically feasible, particularly for those objects possessing a high spatial symmetry. Despite the appearance of frequency shift to adjacent modes that was caused by the slight misalignments and influence of atmospheric turbulence, we can still observe a clearly distinguishable peak of frequency shifts at the desired OAM detection modes, which is associated with the object¡¯s rotational symmetry. The effect of the energy spread accompanied with the transfer of frequency shifts from the dominant OAM modes to their adjacent ones was carefully analysed, which might offer a important value of reference for further exploration of this field. The natural extension of our scheme is to implement the free-space link in a longer distance to detect any rotating bodies \cite{krenn2014communication,krenn2016twisted,tamburini2011twisting}. Moreover, the ability of our scheme to work at the photon-counting regimes suggests the potential to combine the rotational Doppler effect with quantum entangled light source for long-distance entanglement-enhanced remote sensing technique \cite{krenn2015twisted}. In addition, our feasibility study with extremely low light intensities may pave the way towards applications, such as in covert imaging and biological sensing, where a low photon flux is essential as a high photon flux might have detrimental effects \cite{morris2015imaging}.

\section{ACKNOWLEDGMENTS}
This work is supported by the National Natural Science Foundation of China (NSFC) (11474238, 91636109), the Fundamental Research Funds for the Central Universities at Xiamen University (20720160040), the Natural Science Foundation of Fujian Province of China for Distinguished Young Scientists(2015J06002), and the program for New Century Excellent Talents in University of China (NCET-13-0495). R.F. is thankful for financial support by the Banting postdoctoral fellowship of the Natural Sciences and Engineering Research Council of Canada (NSERC).

\bibliographystyle{apsrev4-1}
\bibliography{PRapplied}

\end{document}